\documentclass[twocolumn,letterpaper,showpacs,floatfix,aps]{revtex4}

\usepackage[dvips]{graphics}

\def \beq {\begin{equation}}
\def \eeq {\end{equation}}

\begin{document}

\title{Semi-analytical approach for the Vaidya metric in double-null 
coordinates}

\author{Fernando Girotto}
\affiliation{
Instituto de F\'\i sica, Universidade de S\~ao Paulo,
CP 66318, 05315-970 S\~ao Paulo, SP, Brazil}

\author{Alberto Saa}
\email{asaa@ime.unicamp.br}
\affiliation{
Departamento de Matem\'atica Aplicada,
IMECC -- UNICAMP,
C.P. 6065, 13083-859 Campinas, SP, Brazil.}


\begin{abstract}
We reexamine here a problem considered in detail before by
Waugh and Lake: the solution of spherically symmetric
Einstein's equations with
a radial flow of unpolarized radiation
(the Vaidya metric) in double-null coordinates. This problem is
known to be not analytically solvable, the only known
explicit solutions correspond to the constant mass case
(Schwarzschild solution in Kruskal-Szekeres form) and the linear
and exponential mass functions originally discovered by Waugh and Lake.
We present here a semi-analytical approach that can be used
to discuss some qualitative and quantitative aspects of the Vaidya
metric in double-null coordinates for generic mass functions.
We present also a new analytical solution corresponding to
$(1/v)$-mass function and discuss some physical examples.

\end{abstract}

\maketitle

\section{Introduction}

The Vaidya metric\cite{Kramer}, 
which in radiation coordinates $(w,r,\theta,\phi)$
has the form
\beq
\label{Vaidya}
ds^2 = -\left(1-\frac{2m(w)}{r}\right)dw^2+2cdrdw + r^2d\Omega^2,
\eeq
where $d\Omega^2 = d\theta^2 + \sin^2\theta d\phi^2$, 
$c=\pm 1$, is a solution
of Einstein's equations with spherical symmetry in the eikonal
approximation to a radial flow of unpolarized 
radiation. For the case of an ingoing radial flow, $c=1$ and
$m(w)$ is a monotone increasing mass function in the advanced
time $w$, while $c=-1$ corresponds to an outgoing radial flow,
with $m(w)$ being in this case 
a monotone decreasing mass function in the retarded
time $w$. For many years,   
it has been used in the analysis of spherical collapse and the
formation of naked singularities (For references, see
the extensive list of \cite{Lake} and also \cite{Joshi}).
It is known that Vaidya metric can be obtained, by taking appropriate
limits in the self-similar case, from the Tolman metric representing
spherically symmetric dust distribution\cite{LemosHellaby}. This result has
shed some light on the nature of the so-called shell-focusing
singularities\cite{EardleySmarr}, discussed in details
in \cite{Lake,Kuroda1,LakeZannias1,LakeZannias2}. The Vaidya metric has also proved to 
be useful in the study of Hawking radiation
and the process of black-hole 
evaporation\cite{Hiscock,Kuroda2,Kuroda3,Biernacki,Parentani}, and
more  recently,
in the stochastic gravity program\cite{Bei-Lok}.

Motivated by the longly known fact that the radiation coordinates
are defective at the horizon\cite{LSM}, implying that
the Vaidya metric (\ref{Vaidya}) is not geodesically complete
(see \cite{Fayos} for considerations about possible analytical
extensions),
 Waugh and Lake\cite{WL} 
considered the problem of casting the Vaidya metric in double-null coordinates
$(u,v,\theta,\phi)$.
As all previous attempts to construct a general transformation from
radiation to double-null coordinates have failed, they followed 
Synge\cite{Synge} and considered Einstein's equation with
spherical symmetry in double-null coordinates {\em ab initio}.
The spherically symmetric line element in double-null coordinates is
\beq
\label{uv}
ds^2 = -2f(u,v)du\,dv + r^2(u,v)d\Omega^2,
\eeq
where $f(u,v)$ and $r(u,v)$ are non vanishing functions.
The energy-momentum tensor
of a unidirectional radial flow of unpolarized radiation  
in the eikonal approximation is 
given by
\beq
\label{T}
T_{ab} = \frac{1}{8\pi}h(u,v)k_a k_b,
\eeq
where $k_a$ is a radial null vector. As Waugh and Lake,
we will consider, without loss
of generality, the case of the flow along $v$-direction: 
$k_a = (0,1,0,0)$. The case of simultaneous ingoing and outgoing
flows was considered in \cite{Let}. Einstein's equations are less
constrained in such case, allowing the construction of
 some similarity solutions. In our case,
the metric (\ref{uv}) with the energy-momentum
tensor (\ref{T}) reduce to the following set of equations\cite{WL}:
\begin{eqnarray}
\label{ef}
f(u,v) &=& 2B(v)\partial_u r(u,v), \\
\label{er}
\partial_v r(u,v) &=& -B(v)\left( 1-\frac{2A(v)}{r(u,v)}\right), \\
\label{eh}
h(u,v) &=& -4\frac{B(v) A'(v)}{r^2(u,v)},
\end{eqnarray}
where $B(v)$ and $A(v)$ are arbitrary functions obeying, according
to the weak energy
condition,   
\beq
\label{WE}
B(v)A'(v) \le 0.
\eeq
Note that from (\ref{ef}), the regularity of $f(u,v)$ 
requires $B(v)\ne 0$. 
Schwarzschild solution in the Kruskal-Szekeres form corresponds to 
$A'=0$.
For $A'\ne 0$, the choice
\beq
2B(v) = - \frac{A'}{|A'|}
\eeq
allows one to interpret $A(v)$ as the mass of the solution and
$v$ as the proper time as measured in the rest frame at infinity
for the asymptotically flat case\cite{WL}. We assume here 
$A(v)>0$.
The radial flow is ingoing if $A'>0$, and outgoing if $A'<0$.
Note that if the weak energy condition (\ref{WE}) holds, 
the function $A(v)$ is monotone, implying that 
the radial flow must be ingoing or outgoing for all $v$. 
It is not possible, for instance, 
to have ``oscillating'' mass functions $A(v)$ in double null
coordinates. 

The problem may be stated in the following way: given the
functions $A(v)$ and $B(v)$, 
Eq. (\ref{er}) shall be integrated and $r(u,v)$
is obtained. Then,  
$f(u,v)$ and $h(u,v)$ are calculated from (\ref{ef}) and
(\ref{eh}). The arbitrary function  of $u$ 
appearing in the integration of (\ref{er}) must be chosen properly to have
a non-vanishing $f(u,v)$ function from (\ref{ef}). 
Unfortunately, this procedure is not analytically
solvable in general.
Waugh and Lake, nevertheless, were able to find regular solutions for
Eqs. (\ref{ef})-(\ref{eh}) for linear ($A(v)=\lambda c v$)
and a certain exponential ($A(v)=\frac{1}{\beta}\left(
\alpha\exp(\beta c v/2) + 1 \right) $) 
mass functions ($\lambda,\alpha$, and $\beta$ are 
positive constants, $c=\pm 1$, corresponding to ingoing/outgoing
flow, respectively.). To the best of our knowledge, these are the only 
varying mass analytical solutions obtained in 
double-null coordinates so far. We notice, however, that Kuroda 
was able to construct a transformation from radiation to double-null
coordinates for some particular mass functions\cite{Kuroda2,Kuroda3}.

\begin{figure}[ht]
\resizebox{\linewidth}{!}{\includegraphics*{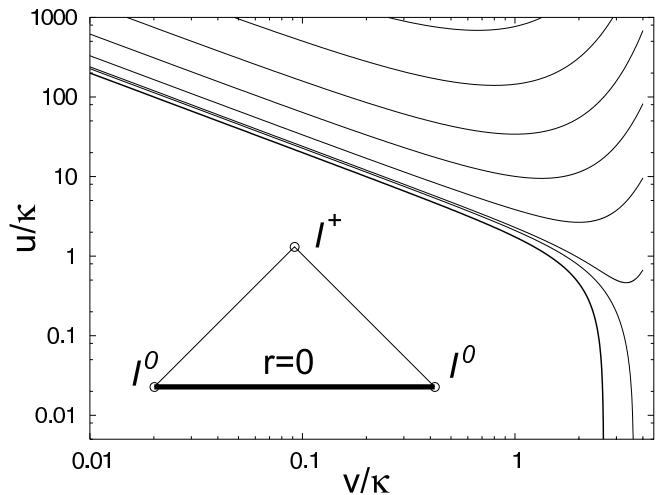}}
\caption{Constant $r/\kappa$ curves for the case $A(v)=\kappa^2/v$. 
The values of $r/\kappa$ shown are, from bottom to top,
$0,0.5, 0.6, 1, 1.5, 2, 2.5,$ and $3$. Future is to right and up.
The corresponding conformal diagram is inserted. Here, $r=0$ is a naked singularity.}
\label{fig1}
\end{figure}

In the following section, we will present a semi-analytical approach
to attack the problem of solving Eqs. (\ref{ef})-(\ref{eh}) for 
general mass functions obeying the weak energy condition (\ref{WE}).
The approach allows us to construct qualitatively conformal diagrams,
identifying horizons and singularities, 
and also to evaluate specific geometric quantities.
Before, however, we notice that one
can solve analytically Eqs. (\ref{ef})-(\ref{eh}) also for the
case of
\beq
\label{inv}
A(v) = \frac{\kappa^2}{v},
\eeq
being $\kappa$ a massive parameter, 
and $B=1/2$.
With the mass function (\ref{inv}), Eq. (\ref{er}) can be integrated
as
\beq
\label{L}
L(r,u,v) = P(u)v+2\kappa\,{e^{\xi^2}}-
\sqrt{2}\,v\int_0^\xi e^{s^2} ds=0,
\eeq
where 
\beq 
\xi=\frac{1}{2\sqrt{2}}\left( \frac{v}{\kappa} + 2 \frac{r}{\kappa} 
\right)
\eeq 
and $P(u)$ is an arbitrary function. It is quite easy
to show that $L(r,u,v)=0$ is an invariant surface of (\ref{er}),
{\em i.e}, one has
\beq
\frac{dL}{dv} = \frac{\partial L}{\partial r}\partial_v r +
\frac{\partial L}{\partial v} = \frac{1}{v}L
\eeq
along the solutions of (\ref{er}). From (\ref{ef}), we have
\beq
f = -\kappa P'(u) v \frac{e^{-\xi^2}}{2r}
\eeq
The function $P(u)$ must the chosen to preclude zeros of $f(u,v)$. 
With $P(u)= -u/\kappa$, we have from (\ref{L})
\beq
\label{r}
\frac{u}{\kappa}\frac{v}{\kappa} = 2\,{e^{\xi^2}}-
\sqrt{2}\,\frac{v}{\kappa}\int_0^\xi e^{s^2} ds. 
\eeq
Equation (\ref{r}) defines $r(u,v)$. Its graphics and  the 
corresponding conformal
diagram is presented in Fig. \ref{fig1}.

The only spacetime singularity for $v>0$ is $r=0$. We recall that he 
Kretschmann scalar for metrics of the form (\ref{uv})
obeying (\ref{ef})-(\ref{eh}) 
is given by\cite{WL}
\beq
K = R_{abcd} R^{abcd} = 48 \frac{A^2(v)}{r^6}.
\eeq
The resulting spacetime corresponds to a naked singularity that
vanishes smoothly, giving rise to an empty spacetime. Its causal
structure is identical to the first linear case 
($\lambda>1/16$)
of Waugh and Lake, with time reversed (See Fig. \ref{fig3} below).

We finish this section noticing that algebraic manipulation programs
(Maple, for instance)
are able to find solutions of (\ref{er})
for the case $A(v)\propto v^2$ in terms
of Airy functions, but they seem too cumbersome to be useful.
Furthermore, their causal structure is always similar to the  
($\lambda>1/16$) linear case
of Waugh and Lake. Also, Kuroda has considered before the case of 
$M(w)=\mu w^n$ for small $w$ and $n\ge 1$ in radiation coordinates,
obtaining some local properties of the singularity 
$r=w=0$\cite{Kuroda3}.

\section{Semi-analytical approach}

Eq. (\ref{er}) along $u$ constant is a first order ordinary differential
equation in $v$. One can evaluate the function $r(u,v)$ in
any point by solving the $v$-initial value problem knowing
$r(u,0)$. The trivial example of Minkowski spacetime ($A=0$),
for instance, can be obtained by choosing $r(u,0)=u/2$.

\begin{figure}[ht]
\resizebox{\linewidth}{!}{\includegraphics*{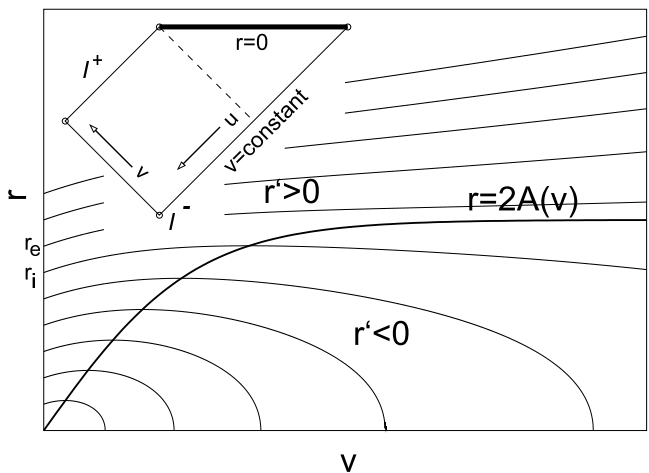}}
\caption{In the region below the generic monotonic function 
$r=2A(v)$ (the apparent horizon), all solutions
have $r'<0$. Any solution that enters into this region will reach
the singularity at $r=0$ with finite $v$. In the other hand, solutions
confined to the $r'>0$ region always escape from the singularity and
reach ${\cal I^+}$. In this case, 
supposing that the mass function $A(v)$ has an
asymptotic value such that $r_{\rm i}(v) < 2A(v) < r_{\rm i}(v)$,
there 
exists an event horizon (the dashed line on the conformal
diagram) somewhere between the solutions $r_{\rm i}(v)$ and $r_{\rm e}(v)$.
The vector $T=\partial_v - \partial_u$ points to the future for the
ingoing radiation case.}
\label{fig2}
\end{figure}

The curve $r=2A(v)$ is the frontier of two regions of the
$(v,r)$ plane where the solutions of (\ref{er}) have qualitative
distinct behaviors. Suppose, for instance, $A'(v)>0$ (and
$B=-1/2$. The case of an outgoing radiation flow follows
in a straightforward manner). For all points of the plane $(v,r)$ below this
curve, $r' < 0$ (See Fig. \ref{fig2}). 
Hence, any solution entering in this region
will, unavoidably, reach the singularity at $r=0$, with finite
$v$. Suppose a given solution $r_{\rm i}(v)$ with initial condition
$r_{\rm i}(0)=r_{\rm i}$ enters into this region. As the uniqueness of
solutions for (\ref{er}) is guaranteed for any point with $r\ne 0$,
solutions never cross each other in the plane $(v,r)$ for $r\ne 0$.
Hence, any solution starting at $r(0)<r_{\rm i}$ is confined the the
region below $r_{\rm i}(v)$, and it will also reach the singularity
at $r=0$ with finite $v$. On the other hand, suppose that
a given solution $r_{\rm e}(v)$ with initial condition
$r_{\rm e}(0)=r_{\rm e}$ never enters into region bellow the curve
$r=2A(v)$. 
Any solution starting at $r(0)>r_{\rm e}$ is, therefore, confined to
the region above $r_{\rm e}(v)$ and will escape from the
singularity at $r=0$. Keeping in mind that these solutions
correspond to null trajectories with $u$ constant, it is possible
to infer the causal structure of the spacetime from their qualitative
behavior  and an appropriate choice
for the initial conditions $r(u,0)$. For instance,
in the case of Fig. \ref{fig2}, supposing that the mass function $A(v)$ has an
asymptotic value such that $r_{\rm i}(v) < 2A(v) < r_{\rm i}(v)$,
an event horizon shall
be located somewhere between $r_{\rm i}(v)$ and $r_{\rm e}(v)$.

Event horizons are global structures. 
One needs to know the total ($v\rightarrow\infty$) spacetime evolution to define them.
On the other hand, apparent horizons are defined locally\cite{HawkingEllis}. Indeed,
  they are the real relevant ones in our semi-analytical approach.
The curve $r=2A(v)$ is an apparent horizon. For a given $v$, all 
solutions in Fig. \ref{fig2} such that $r(v)\le 2A(v)$ will be captured
by the singularity. Solutions for which $r(v)> 2A(v)$ are temporally
free, but they may find themselves trapped later if $A(v)$ increases.
The event horizon is the last of these solutions trapped by the singularity.

We remind that Eq. (\ref{ef}) requires that $\partial_u r(u,v)\ne 0$,
implying that, for $v$ constant, $r(u+du,v)\ne r(u,v)$ for all
$u$. This conditions holds everywhere we have uniqueness of the
solutions of (\ref{er}),
provided that the initial conditions are such that  $\partial_u r(u,0)\ne 0$.
Moreover, if $\partial_u r(u,0)> 0 $, then $\partial_u r(u,v)> 0 $,
implying from (\ref{ef}) that the sign of $f(u,v)$ will be determined by
$B(v)$. For the ingoing radiation case, one has $B(v)=-1/2$, and from (\ref{uv})
one sees that the vector field $T=\partial_v - \partial_u$ is timelike. 
Our conformal diagram are oriented such that this vector field points upward.
If the condition $\partial_u r(u,0)\ne 0 $ holds,
the conformal diagrams obtained for a given $A(v)$ from different initial
conditions are always equivalent.
Some explicit examples will clarify the proposed approach.

\subsection{Linear mass function}

The linear mass function solution discovered by Waugh and Lake
corresponds to (the ingoing radiation case, the outgoing one 
also follows in a straightforward way)
$A = \lambda v$ and 
$B = -1/2$. Note that with these choices, Eq. (\ref{er}) does not
satisfy the Lipschitz condition in $r=0$, implying that
uniqueness is not guaranteed for solutions passing there.
This will be a crucial point to clarify the nature of the 
three qualitative different cases identified by Waugh and Lake, 
corresponding
to $\lambda > 1/16$, $\lambda = 1/16$, and $\lambda < 1/16$.
We notice that 
the linear mass case was also considered in \cite{LakeZannias2} 
in great detail and in a more general situation (the case of a
charged radial null fluid).

The frontier of the region where all the solutions reach the
singularity at $r=0$ is, in this case, the straight line
$r=2\lambda r$.
Taking the $v$-derivative of (\ref{er}), one gets
\beq
r'' = \frac{\lambda}{r}\left(\frac{v}{2r}\left(1
-\frac{2\lambda v}{r}\right)-1 \right).
\eeq
The regions in the plane $(v,r)$ where the solutions obeys
$r''=0$ are the straight lines
\beq
\label{rr}
r = \frac{v}{4} \left(1\pm\sqrt{1-16\lambda} \right).
\eeq
One can reproduce the analysis of Waugh of Lake for the
three qualitative different cases according to the value
of $\lambda$ by considering the possible solutions of (\ref{rr}).
For this purpose, we will consider the solutions of
(\ref{er})
with the initial condition $r(u,0)\propto u$.

\begin{figure}[ht]
\resizebox{\linewidth}{!}{\includegraphics*{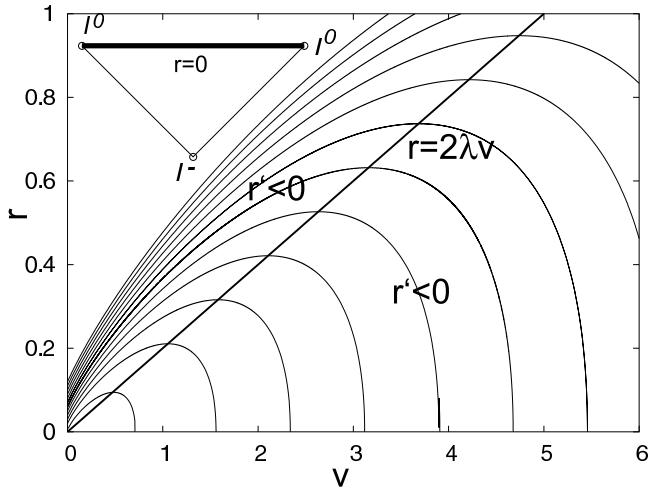}}
\caption{ For $\lambda>1/16$, all solutions are concave for $v>0$.
Hence, all null trajectories along $v$ necessarily reach the
  singularity at $r=0$. The figure corresponds to the case
$\lambda=1/10$. The conformal diagram is inserted.}
\label{fig3}
\end{figure}

For $\lambda>1/16$, the case
showed in Fig. (\ref{fig3}),
$r''<0$ for all points with $v>0$, and the only
relevant frontier is the $r=2\lambda v$ straight line
(the apparent horizon). All solutions
of (\ref{er}) are concave functions and cross the $r'=0$ 
line, reaching the singularity with a finite $v$ which increases
monotonically with $u$.
The causal structure of the corresponding spacetime is very
simple. There is no horizon, and the future cone of all points
ends in the   singularity at $r=0$.

Each curve in Fig \ref{fig3} corresponds to a constant $u$ slice of the
the full solution $r(u,v)$. It is possible, in principle, to
reconstruct the 2-dimensional surface $r(u,v)$ and to plot 
its lines of constant
$r$ on the $(u,v)$ plane, as done in the Fig. (\ref{fig1}),
reproducing all original
results of Waugh and Lake\cite{WL}. However, 
these lines are not necessary to construct the conformal
diagrams.

\begin{figure}[ht]
\resizebox{\linewidth}{!}{\includegraphics*{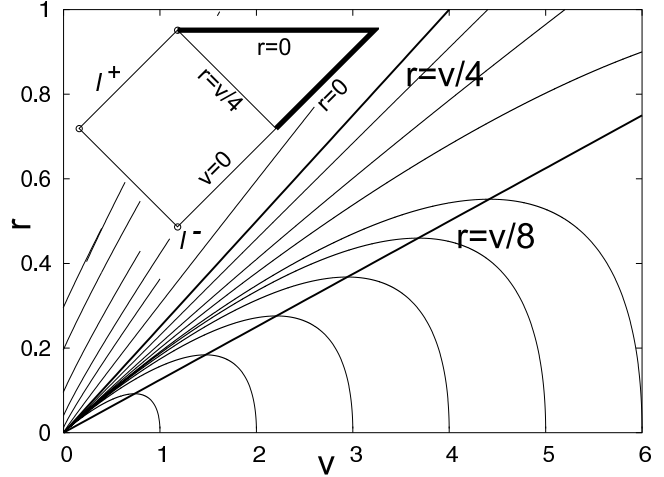}}
\caption{For $\lambda=1/16$, we have two relevant lines 
in the $(r,v)$ plane: $r=v/4$ and $r=v/8$. The first one plays
the role of an event horizon, separating the two regions of distinct
qualitative behavior for the null trajectories along $v$. Beyond
the horizon, all solutions escape to infinity.
Inside, all solutions  reach the singularity 
$r=0$ with finite $v$. The conformal diagram is inserted.
The singularity at $r=0$ is a shell focusing one.}
\label{fig4}
\end{figure}

For $\lambda=1/16$, $r''=0$ on the line $r=v/4$. 
This straight line
itself is a solution. All other solutions are concave functions.
We have two distinct qualitative behavior
for the null trajectories along $u$ constant. All solutions
starting at $r(0)>0$ are confined to the region
above $r=v/4$. They never reach the singularity, all
trajectories reach $\cal I^+$. However, in the region below $r=v/4$, 
we have infinitely many concave trajectories starting and ending
in the (shell-focusing) singularity. They 
 start at $r(0)=0$,
increase in the region between  $r=v/4$ and $r=v/8$, cross
the last line and reach unavoidably $r=0$ again, with finite $v$.
The trajectory $r=v/4$ plays the role of an event
 horizon, separating
two regions with distinct qualitative behavior: one where
constant $u$ null trajectories reach $\cal I^+$ and another
where they start and end in the singularity. This situation
is showed in the Fig. \ref{fig4}. This behavior is only possible,
of course, because the solutions of (\ref{er}) fail to be unique
at $r=0$.

\begin{figure}[ht]
\resizebox{\linewidth}{!}{\includegraphics*{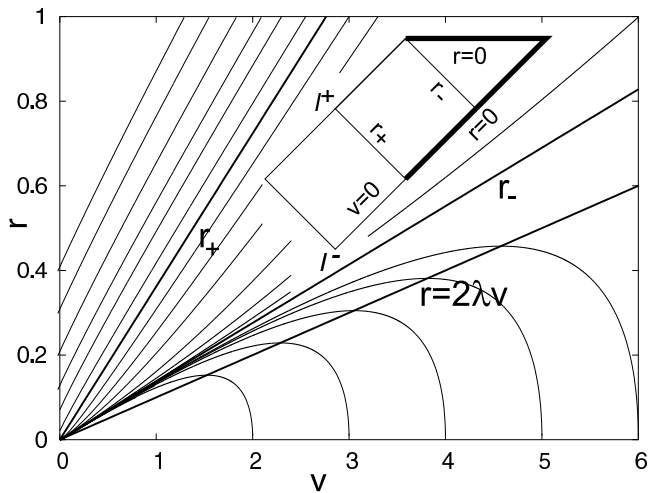}}
\caption{ With $\lambda<1/16$, we have two horizons $r_-$ and $r_+$.
Any solution starting above $r_+$ always escapes to ${\cal I_+}$.
Solutions starting inside the inner event horizon $r_i$ always
reach the singularity at $r=0$. Between the horizons, solutions
starts in the naked singularity but escapes to  $\cal I_+$.
The figure corresponds to the case
$\lambda=1/20$.
The conformal
diagram is inserted. }
\label{fig5}
\end{figure}

For $\lambda<1/16$ (Fig. \ref{fig5}), we have three distinct regions
according to the concavity of the solutions. They are limited by the
two straight lines (\ref{rr}), that are also solutions
of (\ref{er}). Between them, solutions are
convex. Above $r_+$ and below $r_-$, solutions are concave. The last
line is the inner horizon. Inside, the null trajectories with constant $u$
start and end in the singularity $r=0$. The line $r_+$ is an outer
horizon. Between them, the constant $u$ null trajectories starts
in the naked singularity and reach $\cal I^+$. Beyond the outer 
horizon, all  constant $u$ null trajectories escape the singularity.

\subsection{Collapse of a radial null fluid}

\begin{figure}[ht]
\resizebox{\linewidth}{!}{\includegraphics*{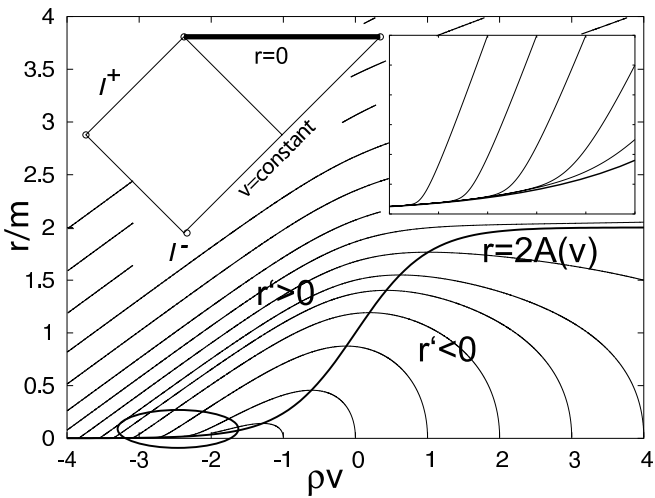}}
\caption{The case $A(v)=m\left( 
1+\tanh\rho v\right)/2$. 
The solutions tend asymptotically to r=0 when $v\rightarrow -\infty$.
A zoom of the range $-3<v<-2$ illustrating such asymptotic behavior
is inserted. In this case, $r=0$ is always singular for 
$v>-\infty$.
This plot corresponds to $m\rho=1$, the situation is
qualitatively identical for other cases. 
The conformal
diagram is inserted. }
\label{fig6}
\end{figure}

As an example of the proposed approach applied to new mass functions,
let us consider first the case of a collapse of incoming radial
null fluid, starting with the empty space. This could be
described, for instance, by the mass function
\beq
\label{collap}
A(v) = \frac{m}{2}(1+\tanh\rho v).
\eeq
The corresponding spacetime is empty for $v\rightarrow -\infty$, it receives
smoothly a radial flow of null fluid, and
finishes as a black hole of mass $m$ for $v\rightarrow \infty$. 
The solutions of (\ref{er}) are presented in the Fig. \ref{fig6}.
With the mass function (\ref{collap}), we always have a singularity
at $r=0$ for $v\ne -\infty$. 
The constant $u$ null trajectories approach $r=0$ asymptotically for
$v\rightarrow -\infty$. Some of them, confined by the horizon,
are captured by the collapse and reach the singularity, and
others scape to $\cal I+$.  Any initial condition obeying
$\partial_u r(u,v_0)\ne 0$ leads to the same causal structure.

The proposed semi-analytical procedure can be used also to derive some
qualitative results. One could, for instance, locate accurately the
event horizon in this case. Since we do not expect deviations from 
the Minkowski spacetime for $v\rightarrow -\infty$, we could
start with initial conditions such that $r(u,-\infty)=u/2$.
Then, we identity in the solutions of (\ref{fig6}) that one
(the horizon $r_{\rm h}(v)$) approaching asymptotically the
curve $r=2A(v)$ for large $v$. For large values of $v_0$, the
horizon would be located at $(2r_{\rm h}(-v_0),-v_0)$.

\subsection{Black hole evaporation}

As a last example, let us consider the physically relevant
case of black hole evaporation. We remind that much work
has been devoted on Vaidya metric to describe black hole
evaporation\cite{Hiscock,Kuroda2,Kuroda3,Biernacki,Parentani}.
Our approach allow us to analyze this question in the more
convenient double-null coordinates. The process of black hole
evaporations supposes $A'<0$ and an outgoing radiation flow.
Hence, one needs to chose $B(v)=1/2$ in our procedure. With this
choice, we see from (\ref{ef}) that $f(u,v)>0$, implying 
that the field $\partial_v-\partial_u$ is not timelike anymore.
It is namely its orthogonal vector field $\partial_v+\partial_u$ that
will define the time evolution in this case. In order to keep
the future direction pointing upward in the conformal diagrams,
one needs to re-orient the coordinates $u$ and $v$. The appropriate
transformation here is
\beq
\label{tr}
v\rightarrow u, \quad u\rightarrow -v.
\eeq
With this choice and $B(v)=1/2$, equation (\ref{er}) reads
\beq
\partial_u r(u,v) = \frac{1}{2}\left( 1-\frac{2A(-u)}{r(u,v)}\right),
\eeq
which is formally identical to the equation for the ingoing case
characterized by the mass function $A(-u)$ with radiation flow
along the $u$ direction.
Hence, the description of an outgoing situation with decreasing
mass function $A(u)$ can be obtained by applying the transformation
(\ref{tr}) for the ingoing case corresponding the mass function $A(-v)$.

\begin{figure}[ht]
\resizebox{0.8\linewidth}{!}{\includegraphics*{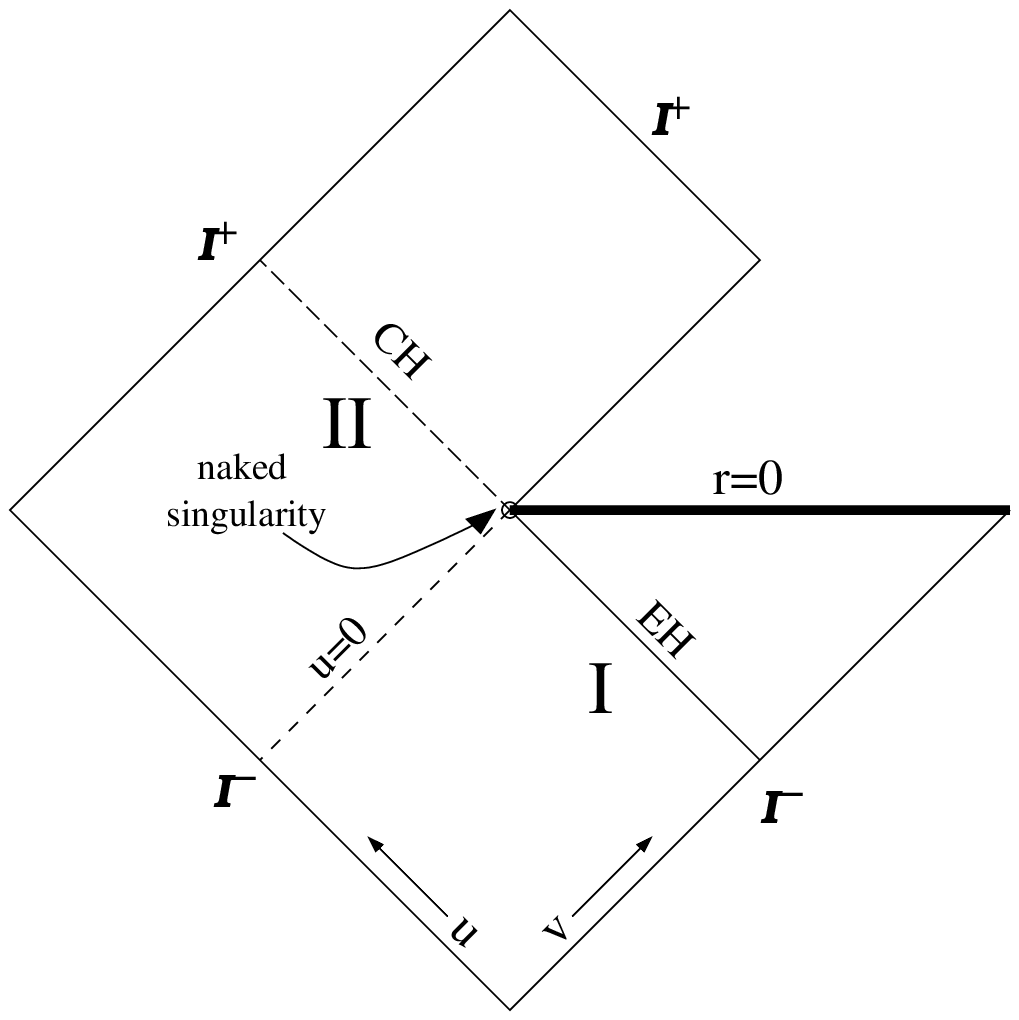}}
\caption{Conformal diagram for a evaporating black hole
with mass $A(u)$ given by (\ref{evap}). The region I corresponds
to the black hole, and can be obtained, by using the transformations
(\ref{tr}), from the ingoing radiation case. Region II is the
empty space left after the vanishing of the black hole at $u=0$.
The event horizon is the line EH. The dashed CH line is a Cauchy
horizon. Beyond it, we have breakdown of predictability due to the
naked singularity left by the black hole. 
}
\label{fig7}
\end{figure}

With the help of (\ref{tr}), one could construct the conformal diagram
corresponding to the mass function $A(u)= m(1-\tanh\rho u)/2$
starting from the last example. Such a case, however, does
not correspond really to an evaporating black hole since, according to
this mass function, the black hole never disappears, the singularity
at $r=0$ is ever present for $u<\infty$. In a real process of
black hole evaporation due to Hawking radiation, the black hole
loses mass with a rate $\dot M \propto M^{-2}$, disappearing completely
in a finite time and giving rise to an empty spacetime.
The mass function
\beq
\label{evap}
A(u) = \left\{\begin{array}{cc} 
-m\tanh \rho u, & u<0, \\
0, & u\ge 0,
\end{array}\right.
\eeq
can be used to simulate the vanishing of a black hole in a finite
time. For the ingoing radiation case $A(-v)$, the situation is similar
to the last example, but with the crucial difference that for $v\le 0$
the spacetime is empty, and the solutions can reach (and cross) 
$r=0$. We notice that the form of $A(u)$ is not important. In order
to construct the conformal diagram of a evaporating black hole, one only
needs that $A(u)=0$ after some $u<\infty$.
In this case, the full conformal diagram (see Fig.  \ref{fig7})
is constructed  adding to that one obtained from the ingoing
collapse case by using the transformation 
(\ref{tr}), a patch of Minkowski spacetime,
just after the instant $u=0$ corresponding to the vanishing of 
the black hole. The resulting conformal diagram represents a
black hole that evaporates leaving behind a naked singularity.

\section{Conclusion}

We presented a semi-analytical approach to construct the Vaidya
metric in double-null coordinates for general monotonic mass functions. 
We applied it to elucidate the nature of the three distinct qualitative
causal structures identified by Waugh and Lake\cite{WL} 
for linear mass functions,
 to construct a solution corresponding to a smooth radial collapse
of a null fluid, starting from empty space and finishing with a
black-hole, and to construct the conformal diagram of a evaporating
black hole.

The approach involves an arbitrary function $r(u,0)$. However,
any choice for which $\partial_u r(u,0)\ne 0$ will result in 
causally equivalent regular
($f(u,v)\ne 0$) 
space-times. The analytical solutions of this
problem also involve an arbitrary function, as $P(u)$
in (\ref{L}), that must be chosen properly in order to get
a regular spacetime.

The double-null form of the Vaidya metric is suitable to
the study of quasi-normal modes of time depending solutions. In
double-null coordinates, 
the equations for scalar perturbations are separable
and we obtain an effective Schr\"odinger equation with a time
depending potential for the perturbations. This situation has been
recently considered in the heuristic analysis of \cite{Hod}.
The full problem is now under investigation.

\acknowledgments

This work was supported by FAPESP and CNPq.

\end{document}